\documentclass[11pt,reqno]{amsart}
\usepackage{amsmath,amssymb,amsfonts,amsthm}

\theoremstyle{plain}
\newtheorem{theorem}{Theorem}
\newtheorem*{lemma}{Lemma}
\newtheorem{corollary}{Corollary}
\newtheorem{proposition}{Proposition}

\theoremstyle{definition}

\newtheorem*{acknowledgements}{Acknowledgements}

\theoremstyle{remark}

\newcommand{\oR}{{\mathbb R}}

\newcommand{\oC}{{\mathbb C}}
\newcommand{\oZ}{{\mathbb Z}}

 \renewcommand{\Im}{\mathop{\mathrm{Im}}\nolimits}
 \newcommand{\eqdef}{\stackrel{\mathrm{def}}{=}}

 \newcommand{\inn}[2]{\left\langle #1\,\mathord{,}\,#2\right\rangle} 

\begin{document}

\title{Star product algebras of test functions}

\author{M.~A.~Soloviev}
\address{Lebedev
Physical Institute, Russian Academy of Sciences, Leninsky Prospect
53, Moscow 119991, Russia} \email{soloviev@lpi.ru}

\thanks{}
\subjclass[2000]{53D55, 46E25, 46F05, 81T05, 46L65}

\keywords{noncommutative quantum field theory,  Moyal star
product, topological $*$-algebra, Gelfand-Shilov space}

 \begin{abstract}
We prove that the Gelfand-Shilov spaces $S^\beta_\alpha$ are
topological algebras under the Moyal $\star$-product if and only
if $\alpha\ge\beta$. These spaces of test functions can be used to
construct a noncommutative quantum  field theory. The star product
depends continuously on the noncommutativity parameter in their
topology. We also prove that the series expansion of the Moyal
product is absolutely convergent in $S^\beta_\alpha$ if and only
if $\beta<1/2$.
 \end{abstract}

\hfill  FIAN/TD/16-07

\hfill Theor. Math. Phys. 153 (2007) 1351-1363
\bigskip


\vspace{2cm}

 \maketitle

\section{Introduction}
\label{S1} In recent years, considerable attention has been given
to noncommutative quantum field theories (QFTs), which  occupy an
intermediate position between the usual QFT and string theory
(see, e.g.,~\cite{Sz} for a review). The interaction terms in the
Lagrangians of these theories are expressed  in terms of a star
product, which is a noncommutative and nonlocal deformation of the
ordinary pointwise product of fields. This deformation leads to
the loss of the commutativity of space-time coordinates and to a
commutation relation of the form
\begin{equation}
[x^\mu, x^\nu]_\star=i\theta^{\mu\nu},
 \label{0.1}
\end{equation}
where  $\theta^{\mu\nu}$ is a real antisymmetric matrix, constant
in the simplest case.

The conceptual framework  of quantum physics on a noncommutative
space-time is  still not conclusively established, and a serious
effort is being made to clarify the questions of the causality of
observables and of the implementation of symmetries, and also of
the conditions of unitarity. In parallel with the study of actual
models, there have also  been attempts~\cite{Alv, Ch,FW}  to
extend the axiomatic approach~\cite{SW,BLOT} to noncommutative
QFT. The description of quantum fields in terms of operator-valued
distributions is one of the cornerstones of the axiomatic
approach, and this raises the question of the optimal choice of
test functions in noncommutative QFT. The relevance of such a
question to finding solutions of QFTs was discussed in~\cite{W}.
There is some evidence that the Schwartz space $S$ used in the
standard formalism~\cite{SW,BLOT} is not quite adequate for a QFT
on  noncommutative space-time. As noted in~\cite{Alv}, the
tempered character of the Schwartz distributions can be
incompatible  with severe singularities  caused by the UV/IR
mixing intrinsic in noncommutative field theories. A further
indication is  an exponential growth of the correlation functions
of some gauge-invariant operators in momentum space, found
in~\cite{IIKK, GHI}. Moreover, the very structure of the star
product, which is defined by an infinite-order differential
operator, suggests that analytic test functions are best suited
for use in a noncommutative QFT. Some subtleties in the derivation
of the CPT and spin-statistics theorems in the enlarged formalism
with analytic test functions were discussed in~\cite{S06}. Here,
we take a different approach to the question  and propose a
criterion for choosing  a suitable test function space; this
criterion implies that this space must be an algebra under the
star product. The analysis is performed in the framework of
Gelfand-Shilov spaces $S^\beta_\alpha$, which are subalgebras of
the Schwartz space with respect to the ordinary product. The index
$\alpha$ determines the behavior of the test functions at
infinity, and $\beta$ determines their smoothness. The smaller
these indices are, the smaller  the space $S^\beta_\alpha$ and the
larger  its dual space of generalized functions. The Schwartz
space $S$ is the formal limit of $S^\beta_\alpha$ as
$\alpha,\beta\to+\infty$.

In Sec.~\ref{S2}, we propose a simple way to analyze two
well-known associative noncommutative products on the Schwartz
space $S(\oR^d)$. Both these products are generated by a Poisson
structure on $\oR^d$. The first product is a noncommutative
deformation of the ordinary pointwise product of functions; its
formal power series expansion in the noncommutativity parameter
reproduces the Weyl-Groenewold-Moyal star product (which is
hereafter called the Moyal product, as in  most  papers on this
subject). We give an example that clearly demonstrates that this
expansion does not converge in general  in the topology of the
Schwartz space. The second product, called the twisted
convolution, is obtained from the first product by the Fourier
transformation. In Sec.~\ref{S3}, we show that the proposed
approach is also applicable to other spaces of test functions.
Specifically, it allows proving that the subspaces
$S^\beta_\alpha\subset S$ remain subalgebras of the Schwartz space
under the noncommutative deformation  if and only if
$\alpha\ge\beta$. In Sec.~\ref{S4}, we study the conditions under
which  the series defining the Moyal $\star$-product converges,
and we show that these conditions result in additional
restrictions on the index $\beta$. In Sec.~\ref{S5}, we prove that
for any $\star$-algebra $S^\beta_\alpha$, the star product depends
on the noncommutativity parameter $\theta$ continuously (i.e.,
this product is indeed a deformation of the ordinary product). We
briefly  discuss the obtained results in Sec.~\ref{S6}.  In the
appendix, we prove an elementary lemma which shows that  the
spaces under consideration contain functions with certain
properties useful in analyzing the operation of star
multiplication in these spaces and in finding the conditions for
the convergence of the power series expansion  in $\theta$ of the
star product.

\section{Star-product structure on the Schwartz space}
\label{S2}
 Let $f$ and $g$ be smooth complex-valued functions on
$\oR^d$, and let $\theta^{\mu\nu}$ be a constant antisymmetric
(possibly degenerate) $d\times d$ matrix. Then the Moyal
$\star_\theta$-product of $f$ and $g$ is defined by the formula
\begin{multline}
(f\star_\theta g)(x)=f(x)e^{i(\overleftarrow{\partial_\mu}\,
\theta^{\mu\nu}\,\overrightarrow{\partial_\nu})/2}g(x)=\\
=f(x)g(x)+\sum_{n=1}^\infty\left(\frac{i}{2}\right)^n\frac{1}{n!}\,\theta^{\mu_1\nu_1}\dots
\theta^{\mu_n\nu_n}\partial_{\mu_1}\dots
\partial_{\mu_n}f(x)\partial_{\nu_1}\dots\partial_{\nu_n}g(x)
 \label{*}
\end{multline}
(summation over the indices  $\mu_1,\dots \mu_n$ and  $\nu_1,\dots
\nu_n$ is implied), which is usually understood as a formal power
series in $\theta$.  Product~\eqref{*} reduces to the ordinary
pointwise product of  functions as $\theta\to 0$. The
order-$\theta$ part coincides with $(i/2)\{f,g\}$, where the
Poisson bracket $\{\cdot,\cdot\}$ is determined by the matrix
$\theta^{\mu\nu}$. In particular,
$$
x^\mu\star_\theta x^\nu-x^\nu\star_\theta x^\mu=i\theta^{\mu\nu}
 $$
and we obtain commutation relation~\eqref{0.1}. Thus,  $\theta$
plays the role of a noncommutativity parameter. Because this
parameter is the same throughout the paper,  we write  $\star$
instead of $\star_\theta$ in what follows.

We now suppose that the functions $f$ and $g$ decrease rapidly at
infinity and belong to the Schwartz space $S(\oR^d)$. Then each
term in series expansion~\eqref{*} has a Fourier
transform,\footnote{We use the  definition of the Fourier operator
$(\mathcal{F}f)(p)=\hat f(p)=\int f(x)e^{-i\inn{p}{x}}dx$. The
bracket $\inn{\cdot}{\cdot}$ denoting pairing of the space $\oR^d$
and its dual $\oR^{d\,\prime}$ is identified with the standard
Euclidean structure on $\oR^d$.} which is readily calculated from
the formulas $(\widehat{\partial_\mu f})(p)=ip_\mu\hat f(p)$ and
$(\widehat{f g})(p)=(2\pi)^{-d}\int\hat f(q)\hat g(p-q)dq$.
Summing  over $n$, we obtain
\begin{equation}
 (2\pi)^{-d}\int \hat
f(q)\hat g(p-q)e^{-(i/2)\theta^{\mu\nu}q_\mu(p_\nu- q_\nu)}dq=
(2\pi)^{-d}\int \hat f(q)\hat g(p-q)e^{(i/2)\theta^{\mu\nu}p_\mu
q_\nu}dq.
 \label{1.1}
\end{equation}
In what follows, we  use the index-free notation  (whenever
possible). The Poisson tensor $\theta^{\mu\nu}$ determines the
antisymmetric bilinear form $\theta^{\mu\nu}p_\mu q_\nu$ on
$\oR^{d\,\prime}\times\oR^{d\,\prime}$ and can be identified with
the operator $\theta\colon \oR^{d\,\prime}\to \oR^d$ which takes
each  element $p\in \oR^{d\,\prime}$ with the coordinates $p_\nu$
to a vector with the coordinates $(1/2)\theta^{\mu\nu}p_\nu$. (The
coefficient $1/2$ is inserted to simplify the formulas that
follow.) The function
\begin{equation}
(\hat f\circledast\hat g)(p) \eqdef \int \hat f(q)\hat
g(p-q)e^{i\inn{p}{\theta q}}dq
 \label{1.2}
\end{equation}
is called the twisted convolution of $\hat f$ and $\hat g$. We let
$\tau_p$ denote the shift operator $\hat f(q)\to \hat f(q-p)$ and
$\tau_-$ the reflection  $\hat f(q)\to \hat f(-q)$. Then the
twisted convolution is obtained from the ordinary convolution
$\int\hat f(q)(\tau_p\tau_- \hat g)(q)\,dq$ by replacing $\tau_p$
with the operators $e^{i\inn{p}{\theta q}}\tau_p\,$, which
implement a projective representation of the translation group.
Since $\hat f,\hat g\in S=\mathcal{F}[S]$, it can be easily seen
that function~\eqref{1.2} is smooth and rapidly decreasing, i. e.,
it also belongs to the Schwartz space. The binary operation $(\hat
f, \hat g)\to \hat f \circledast\hat g$ is associative. Indeed, we
have
\begin{multline*}
((\hat f\circledast \hat g)\circledast \hat h)(p)= \int\left\{\int
\hat f(k)\hat g(q-k)e^{i\inn{q}{\theta k}}dk
\right\}\hat h(p-q)e^{i\inn{p}{\theta q}}dq\\
=\int \hat f(k)\left\{\int\hat g(q)\hat
h(p-k-q)e^{i\inn{p-k}{\theta q}}dq \right\} e^{i\inn{p}{\theta k}}
dk =(\hat f\circledast (\hat g\circledast \hat h))(p).
\end{multline*}
Obviously, $(\hat f\circledast \hat g)^*=\hat g^*\circledast \hat
f^*$. Therefore $(S,\circledast)$ is an involutive algebra with
the complex conjugation as involution.

We  let $f\times g$ denote  the element of $S$ whose Fourier
transform is function~\eqref{1.1}, i.e.,
\begin{equation}
\widehat{f\times g} =(2\pi)^{-d}\,\hat f\circledast\hat g,\qquad
f,g\in S(\oR^d).
 \label{1.3}
\end{equation}
More explicitly,
 \begin{equation}
(f\times g)(x)=\frac{1}{(2\pi)^{2d}}\int\int \hat f(q)\hat
g(p)\,e^{i\inn{q}{x}+i\inn{p}{x}-i\inn{q}{\theta p}}dq dp.
\label{1.4}
\end{equation}
This function is  called the twisted product of $f$ and $g$. If
the matrix $\theta$ is invertible and  the Poisson structure is
hence symplectic, then
 \begin{equation}
(f\times g)(x) =\frac{1}{(2\pi)^d|\det \theta|}\int\int
      f(y)g(z)\,e^{i\inn{\theta^{-1}(x-y)}{x-z}}dydz.
 \label{1.5}
 \end{equation}
It is easily seen that nonlocal product~\eqref{1.5} is translation
and symplectic equivariant,\footnote{It can be shown
that~\eqref{1.5} follows from these properties combined with
associativity and nonlocality.} as is the ordinary pointwise
product. Applying the inverse Fourier transformation to the power
series expansion in $\theta$ of $(2\pi)^{-d}\,\hat
f\circledast\hat g$, we obtain precisely  initial
series~\eqref{*}. But it  cannot be asserted that this series
converges to $f\times g$ in the Schwartz space whose topology is
determined by the norms\footnote{In \eqref{1.6}, $\kappa\in
\oZ_+^d$, and the notation $|\kappa|=\kappa_1+\cdots+\kappa_d$,
$\partial^\kappa=\partial^{|\kappa|}/(\partial
x_1^{\kappa_1}\cdots\partial x_d^{\kappa_d})$ is used.}
\begin{equation}
\|f \|_N=\sup_{x\in\oR^d}\sup_{|\kappa|\leq
N}\,(1+|x|)^N|\partial^\kappa f(x)|.
 \label{1.6}
\end{equation}
This is evident from the simplest example of Gaussian functions.

 \begin{proposition}\label{P1}
  Let $d=2$  and suppose that $\theta^{\mu\nu}=
 \left(\begin{matrix} 0&1\\-1&0\end{matrix}\right)$.
  Let $f(x)=e^{-\gamma|x|^2}$, where $\gamma>1$.
 Then the series expansion of $f\star f$ given by~\eqref{*} does
 not converges in the topology of $S(\oR^2)$.
\end{proposition}

 \begin{proof} We consider the linear functional  $u$ on $S(\oR^2)$,
defined by  $u(f)=\int f(0,x_2)\,dx_2$. Clearly, it is continuous
in  the topology of $S(\oR^2)$, because $|u(f)|\le C \|f\|_2$,
where $C=\int (1+|x_2|)^{-2}dx_2$. Let the terms in
series~\eqref{*} be denoted $h_n$. Then
 $$
u(h_n)=\int h_n(0,x_2)dx_2= \frac{1}{2\pi}\int\hat h_n(p_1,0)dp_1,
 $$
 where
 \begin{equation}
 \hat h_n(p)=\frac{i^n}{(2\pi)^2n!}\int \hat f(q)\hat
g(p-q){\inn{p}{\theta q}}^n\,dq.
  \notag
\end{equation}
We show that  if $\hat f(p)=\hat g(p)=
(\pi/\gamma)e^{-|p|^2/4\gamma}$ and $\gamma>1$, then
$u(h_n)\nrightarrow 0$ as $n\to \infty$. Taking into account that
$\inn{p}{\theta q}=(p_1q_2-p_2q_1)/2$ in this case, we obtain
\begin{equation}
u(h_n)=\frac{1}{8\pi\gamma^2}\left(\frac{i}{2}\right)^n
\frac{1}{n!}\int \left\{\int
e^{-(q_1^2+(p_1-q_1)^2)/4\gamma}dq_1\right\}p_1^n\,dp_1 \int
e^{-q_2^2/2\gamma}q_2^n\,dq_2.
 \label{1.7}
\end{equation}
Let $n$ be even. An elementary calculation gives
$$
|u(h_n)|=\sqrt{\frac{\pi}{2\gamma}}\,\,\frac{\gamma^n}{n!}\,[1\cdot3\dots
(2n-1)]^2\ge \sqrt{\frac{\pi}{2\gamma}}\,\,\frac{\gamma^n}{n}.
$$
This proves the proposition.
\end{proof}

Because the Fourier transformation is an automorphism of the
Schwartz space, this space also is an involutive algebra under the
twisted product $\times$. Moreover, both the algebras
$(S,\circledast)$ and $(S,\times)$ are topological. This can be
verified by a straightforward estimation with norms~\eqref{1.6}
(see, e.g., \cite{G-BV}). A different way is based on using
formula~\eqref{1.4}, which shows that the map $(f,g)\to f\times g$
is representable as the composition of five maps
\begin{equation}
S(\oR^d)\times
S(\oR^d)\stackrel{\otimes}{\longrightarrow}S(\oR^{2d})
\stackrel{\mathcal F}{\longrightarrow}S(\oR^{2d})\stackrel{\cdot
e^{-i(q,\theta p)}}{\longrightarrow} S(\oR^{2d})\stackrel{\mathcal
F^{-1}}{\longrightarrow} S(\oR^{2d})\stackrel{\widehat {\mathsf
m}}{\longrightarrow}S(\oR^d),
 \label{1.9}
\end{equation}
where  $\widehat {\mathsf m}$ denotes the restriction to the
diagonal $h(x,y)\to h(x,x)$. By Schwartz's kernel theorem, the
space $S(\oR^{2d})$ coincides with the completion of the tensor
product $S(\oR^d)\mathbin{\otimes_\pi} S(\oR^d)$ endowed with the
projective topology. The map $(f,g)\to f\otimes g$ is continuous
in this topology. Furthermore, there is a one-to-one
correspondence between the set of continuous bilinear maps
$S(\oR^d)\times S(\oR^d)\to S(\oR^d)$ and the set of continuous
linear maps $S(\oR^{2d})\to S(\oR^d)$. In particular, the linear
map $\widehat {\mathsf m}$ is associated with the ordinary
multiplication $\mathsf m\colon (f,g)\to f\cdot g$, and the
continuity of $\widehat {\mathsf m}$ follows from (and amounts to)
the fact that $S(\oR^d)$ is a topological algebra under ordinary
multiplication. The Fourier transformation is not only linear but
also a topological automorphism of $S$, and the function
$e^{-i(q,\theta p)}$ is obviously an multiplier for this space,
i.e., the multiplication by this function maps $S$ into itself
continuously. Therefore, all the maps involved in
composition~\eqref{1.9} are continuous; hence the algebras
$(S,\circledast)$ and  $(S,\times)$ are indeed topological.

Representation~\eqref{1.9}  also allows finding subalgebras of
$(S,\circledast)$ and $(S,\times)$ that become complete
topological algebras when they are endowed  with an appropriate
topology.

\section{The algebras $(S^\beta_\alpha,\times)$ and
$(S_\beta^\alpha,\circledast)$}
\label{S3}

We recall the definition and basic properties of the $S$-type
spaces introduced by Gelfand and Shilov~\cite{GS2}. The space
$S^\beta_\alpha(\oR^d)$, where $\alpha, \beta\ge 0$, consists of
the functions $f\in S$  satisfying the inequalities\footnote{For
$\alpha=0$, the exponential in~\eqref{2.1} should be replaced with
the characteristic function of the set $|x|\leq A$.}
\begin{equation}
|\partial^\kappa f(x)|\le C B^{|\kappa|}
\kappa^{\beta\kappa}e^{-|x/A|^{1/\alpha}},
 \label{2.1}
\end{equation}
where the constants $C$, $A$, and $B$  depend on $f$ and the
conventional multi-index notation is used, in particular,
$\kappa^{\beta\kappa}=\kappa_1^{\beta\kappa_1}\dots
\kappa_d^{\beta\kappa_d}$.  (If $\kappa_i=0$, then
$\kappa_i^{\beta\kappa_i}$ is taken to be~1).  We also write
$S^\beta_\alpha$ for this space when  this cannot lead to
confusion. It is the union of the Banach spaces   $S^{\beta,
B}_{\alpha, A}$ with the norms
\begin{equation}
\|f \|_{A, B}=\sup_{x,\kappa}\,e^{\,|x/A|^{1/\alpha}}\left|\,
\frac{\partial^\kappa
f(x)}{B^{|\kappa|}\kappa^{\kappa\beta}}\right|.
 \label{2.2}
\end{equation}
and is endowed with the inductive limit topology by the natural
maps $S^{\beta, B}_{\alpha, A}\to S^\beta_\alpha$. The space
$S^\beta_\alpha$ is nontrivial if $\alpha+\beta\ge 1$ with the
exceptional cases $\alpha=0$ and  $\beta=0$, where the
nontriviality conditions are the respective strict inequalities
$\beta>1$ and $\alpha>1$. As shown in~\cite{GS2}, the connecting
maps $S^{\beta, B}_{\alpha, A}\to S^{\beta, B'}_{\alpha, A'}$,
$A'>A$, $B'>B$, are compact. Hence, $S^\beta_\alpha$ is a complete
Montel (perfect) space. The Fourier transformation is a linear
topological isomorphism of $S^\beta_\alpha$ onto $S_\beta^\alpha$.
Every nontrivial space of type $S$ is a topological algebra under
 ordinary multiplication as well as under  ordinary convolution.

\begin{theorem}\label{T1}
 If $\alpha\geq \beta$, then $S_\alpha^\beta(\oR^d)$ is a
topological algebra under twisted product~\eqref{1.5} and
$S^\alpha_\beta(\oR^d)= \mathcal F[S_\alpha^\beta (\oR^d)]$ is a
topological algebra under
 twisted convolution~\eqref{1.2}.
\end{theorem}

 \begin{proof}
By~\eqref{1.3}, the second statement in the theorem is equivalent
to the first. As  Mityagin showed~\cite{M}, the spaces of type $S$
are nuclear and
 $$
 S^\beta_\alpha(\oR^d)\mathbin{\widehat\otimes}_\pi S^\beta_\alpha(\oR^d)=
 S^\beta_\alpha(\oR^{2d}),
 $$
 where  the hat on $\otimes$  denotes  completion.
It therefore  suffices to prove that the function
$e^{-i\inn{q}{\theta p}}$ is a multiplier for
$S_\beta^\alpha(\oR^{2d})$ under the indicated restriction on the
indices  $\alpha$ and $\beta$. Then the operation $(f, g)\to
f\times g$ is representable as the composition of  continuous maps
 \begin{equation}
S^\beta_\alpha(\oR^d)\times
S^\beta_\alpha(\oR^d)\stackrel{\otimes}{\longrightarrow}S^\beta_\alpha(\oR^{2d})
\stackrel{\mathcal
F}{\longrightarrow}S_\beta^\alpha(\oR^{2d})\stackrel{\cdot
e^{-i(q,\theta p)}}{\longrightarrow}
S_\beta^\alpha(\oR^{2d})\stackrel{\mathcal
F^{-1}}{\longrightarrow}
S^\beta_\alpha(\oR^{2d})\stackrel{\widehat {\mathsf
m}}{\longrightarrow}S^\beta_\alpha(\oR^d),
 \label{2.4}
\end{equation}
in  complete analogy with  the case of the Schwartz space
considered in  the preceding section.

According to~\cite{GS2}, a function $\chi(s)$ is a multiplier of
$S^\alpha_\beta$ if it satisfies the estimate
\begin{equation}
|\partial^\kappa\chi(s)|\leq C_\epsilon A_\epsilon^{|\kappa|}
\kappa^{\alpha \kappa} \exp\{\,|\epsilon s|^{1/\beta}\}
 \label{mult}
\end{equation}
 for any  $\epsilon>0$.  Here, we have an entire function of
 $2d$ variables and the required estimate can easily be
derived from the Cauchy inequality
\begin{equation}
|\partial^\kappa\chi(s)|\leq \kappa!\, r^{-|\kappa|} \sup_{w\in
D_r}|\chi(s-w)|,
 \label{2.5}
\end{equation}
where  $D_r=\{w\in \oC^{2d} \colon |w_j|< r\, \forall j\}$. We set
$s=(p,q)$ and $w=(u,v)$ and use the notation
$|\theta|=\sum_{j,k}|\theta^{jk}|$. Then
$|\Im\inn{q-v}{\theta(p-u}|\le r\,|\theta|\,(|q|+|p|+2r)$, and we
obtain
\begin{equation}
|\partial^\kappa \exp\{-i\inn{q}{\theta p}\}|\leq
\kappa!\,r^{-|\kappa|} \exp\{\,r\,|\theta|\,(|s|+2r)\},
 \label{2.6}
\end{equation}
where $|s|=|p|+|q|$. Because $\kappa!\le \kappa^\kappa$ and the
radius $r$ of the polydisk can be taken arbitrarily small, we
immediately conclude that the function $e^{-i\inn{q}{\theta p}}$
is a multiplier for $S^1_1$ and also for any space
$S^\alpha_\beta$ with the indices satisfying $\alpha\ge 1$ and
$\beta\le 1$. In particular, this is the case for all $S_0^\alpha$
because they are nontrivial only if $\alpha>1$.

If $\beta>1$, then we take
\begin{equation}
r=\frac{1}{|\theta||s|}|\epsilon s|^{1/\beta}.
 \label{2.7}
\end{equation}
This expression tends to zero as $|s|\to\infty$. If $r$ is chosen
thus, then the exponential in the right-hand side of~\eqref{2.6}
does not exceed $Ce^{|\epsilon s|^{1/\beta}}$ everywhere in the
region $|s|\ge 1$. Furthermore, we have
\begin{equation}
\frac{\kappa!}{r^\kappa}\le A^{|\kappa|}\kappa^{\beta\kappa}
\sup_\kappa\frac{1}{(Ar)^{|\kappa|}\kappa^{(\beta-1)\kappa}}\leq
A^{|\kappa|}\kappa^{\beta\kappa}
e^{(2d\beta/e)|Ar|^{-1/(\beta-1)}}.
 \notag
\end{equation}
 Substituting~\eqref{2.7}, we see that the last exponential is also
 dominated by $e^{|\epsilon s|^{1/\beta}}$ if $A$ is sufficiently large.
Consequently, the function under consideration is a multiplier for
$S^\beta_\beta$,  $\beta>1$,  as well as for all $S^\alpha_\beta$
whose indices satisfy the inequalities $1<\beta\leq\alpha$.

 In the case $1/2\leq\beta<1$, we use the Young inequality
\begin{equation}
ab\leq\beta a^{1/\beta}+(1-\beta) b^{1/(1-\beta)},\qquad a, b\geq
0,
 \label{2.8}
\end{equation}
 setting  $a=|\epsilon s|$ and
$b=r|\theta|/\epsilon$. Choosing $r=|\kappa|^{1-\beta}$, we find
that the right-hand side of~\eqref{2.6} is dominated by the
expression  $A_\epsilon^{|\kappa|}\kappa^{\beta
\kappa}e^{\,|\epsilon s|^{1/\beta}}$. Therefore, in this case,
$e^{-i\inn{q}{\theta p}}$ is a multiplier for $S^\beta_\beta$ and
for $S^\alpha_\beta$, where $\alpha>\beta$. Finally, if
$0<\beta<1/2$, then  we again use~\eqref{2.8}, but we now take
$r=|\kappa|^\beta$ and conclude that $e^{-i\inn{q}{\theta p}}$ is
a multiplier for $S^{1-\beta}_\beta$. This completes the proof
because the spaces $S_\beta^\alpha$ are trivial if
$\alpha<1-\beta$.
\end{proof}

We note that the Fourier-invariant spaces $S^\beta_\beta$ are
topological algebras under  both of the operations $\times$ and
$\circledast$. The space $S^{1/2}_{1/2}$ is smallest of these and
plays a special role.

We  now show  that the restrictions imposed by Theorem~\ref{T1} on
the indices of the spaces of type  $S$  are necessary for these
spaces to be star-product algebras.

\begin{theorem}\label{T2}
Let the  $\times$-product be determined by a nondegenerate matrix
$\theta^{\mu\nu}$. If $\alpha<\beta$ and the space
$S^\beta_\alpha(\oR^d)$ is nontrivial, then this space contains
functions $f$ and $g$ such that $f\times g\notin
S^\beta_\alpha(\oR^d)$.
\end{theorem}

\begin{proof} Because the matrix $\theta^{\mu\nu}$ is antisymmetric,
 definition~\eqref{1.5} can be rewritten as
 \begin{equation}
     (f\times g)(x)=\frac{1}{(2\pi)^d |\det \theta|}
     \int\int f(y)g(z)e^{i\inn{\theta^{-1}(z-y}{x}
     +i\inn{\theta^{-1}y}{z}}dy dz.
 \label{13}
 \end{equation}
 We first consider the simplest case $\alpha=0$.
All elements of $S_0^\beta$ are compactly supported. It is evident
from~\eqref{13} that the $\times$-product of  such functions
admits an analytic continuation to $\oC^d$. But nontrivial
analytic functions cannot have compact support, and we conclude
that the product $f\times g$
 of two elements of $S^\beta_0$ belongs to the same space only if $(f\times
g)(x)\equiv 0$. But we can easily find functions $f,g\in
S^\beta_0$ such that $(f\times g)(0)> 0$. Indeed, we have
\begin{equation}
(f\times g)(0)= \frac{1}{|\det \theta|} \int f(y)\hat
g(-\theta^{-1}y)dy.
\label{14}
\end{equation}
Because $S^\beta_0$ and $S_\beta^0$ are algebras under the
ordinary multiplication, we can construct nonnegative functions
$f\in S^\beta_0$ and $\hat g\in S_\beta^0$ starting from any
nontrivial elements of these spaces. Furthermore, we can make the
integrand in~\eqref{14} nonvanishing by using the translation
invariance of $S^\beta_0$. Then $f\times g\notin S^\beta_0$.

We now suppose  that $0<\alpha<\beta<1$. We take $f, g\in
S^\beta_{1-\beta}$ such that  $(f\times g)(0)> 0$. These functions
decrease no worse than exponentially of order  $1/(1-\beta)$ with
a finite type. Using  Young inequality~\eqref{2.8} with
$b=|\theta^{-1}(z-y)|/B$ and $a=B|\Im x|$, where $B$ is
sufficiently large, we deduce that $(f\times g)(x)$ can be
analytically continued to $\oC^d$ as an entire function of an
order $\leq 1/\beta$. We consider the analytic continuation in the
variable $x^1$ for $x^2=\dots=x^d=0$. It is well known that any
nontrivial entire function of finite order of growth  cannot have
an exponential decrease of a greater order along a direction of
the complex plane. (This is an immediate consequence of
Theorem~2.5.4 in~\cite{E}.) Therefore the inequality
$1/\alpha>1/\beta$ implies that $f\times g\notin S^\beta_\alpha$.
If $0<\alpha<1$ and $\beta>1$, then we obtain the same conclusion
taking functions $f$ and $g$ in $S^{\beta'}_{1-\beta'}$, where
$\alpha<\beta'<1$.

Let $\alpha=1$ and  $\beta>\alpha$. We again take functions $f$
and $g$ in $S^\beta_0$ such that $f\times g\not\equiv 0$. Then the
analytic continuation of the product $f\times g$ is an entire
function of order~1 and finite type. By the Paley-Wiener theorem,
the support of $\widehat{f\times g}$ is compact. We can also
demonstrate this by shifting the plane of integration in
representation~\eqref{1.2}. By the Cauchy-Poincar\'e theorem, this
leaves the integral unchanged because of the analyticity and rapid
degrease of the elements of $S^0_\beta$ at the real infinity.
Namely, for any $u\in \oR^d$, we have the estimates
\begin{equation}
|\hat f(q+iu)|\leq C\,e^{-|q/B|^{1/\beta}+r|u|},\qquad |\hat
g(p-q-iu)|\leq C'\,e^{r'|u|}
 \notag
\end{equation}
and hence
\begin{equation}
(\hat f\circledast\hat g)(p) =\int \hat f(q+iu)\hat
g(p-q-iu)e^{i\inn{p}{\theta q}-(p,\theta u)}dq.
 \notag
\end{equation}
 Assuming for simplicity that $d=2$ and $\theta=
\left(\begin{matrix}0&1\\-1&0\end{matrix}\right)$, we obtain
\begin{equation}
|(\hat f\circledast\hat g)(p)| \leq
C^{\prime\prime}e^{(r+r')|u|-p_1u_2+p_2u_1}.
 \label{15}
\end{equation}
The support of this convolution is therefore contained in the
 square $\max\{|p_1|,|p_2|\}\leq r+r'$. Indeed, we assume  that
$p_1>r+r'$, for example. If $u_1=0$ and $u_2\to+\infty$, then the
right-hand side of~\eqref{15} vanishes. Because all elements of
$S^1_\beta$ are analytic, we deduce that что $f\times g\not\in
S_1^\beta$.

 Considering the remaining case $\alpha>1$, we again assume that  $d=2$
 and $\theta=\left(\begin{matrix} 0&1\\-1&0\end{matrix}\right)$.
 This does not result in any loss of generality, because
 the spaces $S^\beta_\alpha$ are invariant under the linear changes of
 variables, and we can use a symplectic basis in $\oR^d$. Let  $f(x)=f_1(x_1)f_2(x_2)$
 and $g(x)=g_1(x_1)g_2(x_2)$, where $f_i,g_i\in S^\beta_\alpha(\oR)$.
A simple calculation gives
\begin{equation} (f\times
g)(x_1,0)=\frac{1}{2\pi}\int\! \hat f_1(z_2)
g_2(-z_2)e^{ix_1z_2}dz_2\int\! f_2(y_2)\hat
g_1(y_2)e^{ix_1y_2}dy_2.
 \label{16}
\end{equation}
We set $g_1(\xi)=f_1(\xi)$ and $g_2(\xi)=f_2(-\xi)$. Then
$$
(f\times g)(x_1,0)= h^2(x_1),
$$
where $h=\mathcal F^{-1}(\hat f_1 f_2)$. Because $\alpha>1$, the
function $f_1$ can be chosen such that its Fourier transform $\hat
f_1$ is identically equal to 1 in a neighborhood of zero. As shown
in the appendix, the space $S^\beta_\alpha(\oR)$ contains a
function whose successive derivatives are not less than $n^{\beta
n}$ in absolute value. Let $f_2$ be such a function. Then
\begin{equation}
\partial^n (\hat f_1 f_2)(0)=\partial^n \hat h(0)\ge n^{\beta n},
\qquad n=0,1,2,\dots.
 \label{17}
\end{equation}
It follows that $f\times g\not\in S^\beta_\alpha$, because
 the function $h$ would otherwise satisfy the inequality
\begin{equation}
|h(\xi)|\le C e^{-\left|\frac{\xi}{A}\right|^{1/\alpha}}
\label{18}
\end{equation}
with some constants $C, A>0$,  and we would then  have
\begin{multline}
|\partial^n \hat h(0)|\le C\int |\xi|^ne^{-|\xi/A|^{1/\alpha}}d\xi
=2CA^{n+1}\int_0^\infty t^n e^{-t^{1/\alpha}}dt\\
\le C'A^n\max_{t>0}\left(t^n e^{-(1/2)t^{1/\alpha}}\right)=C'
A^n(2\alpha n/e)^{\alpha n}, \notag
\end{multline}
which contradicts  inequality~\eqref{17} for $\beta>\alpha$.
Theorem~2 is thus proved.
\end{proof}

 The following analogue of Proposition~\ref{P1} holds.

 \begin{proposition}
 Let $d=2$, $\theta^{\mu\nu}=\left(\begin{matrix}
 0&1\\-1&0\end{matrix}\right)$, and $\alpha>\beta$. If $\beta\ge 1/2$,
then there is a function $f\in S^\beta_\alpha(\oR^2)$ such that
the series expansion of $f\star f$ given by~\eqref{*} does not
converges in the topology of $S^\beta_\alpha(\oR^2)$.
\end{proposition}

 \begin{proof}
 We consider the same linear functional $u(f)$ as in the proof of
 Proposition~\ref{P1}. Clearly it is continuous in the topology of
 $S^\beta_\alpha(\oR^2)$. Using the lemma proven in the appendix and
 taking into account that the space
$S_\beta^\alpha(\oR)$ is dilation invariant, we see that under the
condition $\beta\ge 1/2$, it contains a positive even function
that dominates the Gaussian function $e^{-|s|^2/(4\gamma)}$.  We
also note that  $\gamma$ can be taken arbitrarily
 large. Let $\hat f(p_1,p_2)$ be the tensor product of such
   functions. We then  have
    $$
    \hat f(p)\ge  e^{-|p|^2/(4\gamma)}.
    $$
The further arguments are similar to those used to prove
Proposition~\ref{P1}, with the only difference that \eqref{1.7} is
replaced with the estimate
 $$
|u(h_n)|\ge \frac{c}{2^n n!}\int \left\{\int
e^{-(q_1^2+(p_1-q_1)^2)/(4\gamma)}dq_1\right\}p_1^n\,dp_1 \int
e^{-2 q_2^2/(2\gamma)}q_2^n\,dq_2,
 \label{1.8}
$$
which holds for every even integer $n$.
\end{proof}

\section{Convergence of the $\star$-product}

\label{S4}
 The next theorem establishes a simple sufficient
condition for the pointwise convergence of the series obtained
from~\eqref{*} by the Fourier transformation.

\begin{theorem}\label{T3}
If $f,g \in S^\beta_\alpha(\oR^d)$, where $\beta<1$, then the
series
\begin{equation}
\sum_n\frac{i^n}{n!}\int \hat f(q)\hat g(p-q){\inn{p}{\theta
q}}^n\,dq
 \label{e}
\end{equation}
 converges to the function $(\hat f\circledast\hat g)(p)$ uniformly
 on every compact set $Q\subset\oR^d$.
\end{theorem}

 \begin{proof} The condition $\beta<1$ implies that
\begin{equation}
|\hat f(q)|\leq C_a e^{-a|q|},\qquad |\hat g(q)|\leq C'_a
e^{-a|q|}
 \label{3.1}
\end{equation}
for any $a>0$. Let $r$ be so large that $Q$ is contained in the
ball $|p|<r$, and let $a>2r|\theta|$. Then for any $N=0,1,\dots$
and $R>0$, the  estimate
\begin{gather}
\left|\sum_{n=0}^N\frac{i^n}{n!}\int_{|q|>R} \hat f(q)\hat
g(p-q){\inn{p}{\theta q}}^n\,dq-\int_{|q|>R} \hat f(q)\hat
g(p-q)e^{i\inn{p}{\theta q}}dq\right|\leq \notag\\
\leq C'_a\int_{|q|>R} |\hat f(q)|\left( e^{|\inn{p}{\theta q}|}+
1\right)\leq\notag \\ \leq C_aC'_a
\int_{|q|>R}e^{-2r|\theta||q|}(e^{r|\theta| |q|}+1)dq,\qquad p\in
Q,
 \label{3.2}
 \end{gather}
holds. We take $R$ so large that the right-hand side
of~\eqref{3.2} is less than $\epsilon/2$. Next, we choose
$N_\epsilon$ such that for $N>N_\epsilon$, the inequality
$$
\sup_{|p|\le r; |q|\le R}\left| \sum_{n=0}^N
\frac{i^n}{n!}{\inn{p}{\theta q}}^n-e^{i\inn{p}{\theta
q}}\right|<\frac{\epsilon}{2v_RC_aC'_a}
$$
holds, where $v_R$ is the volume of the ball $|q|<R$. Then we have
$$
\sup_{p\in Q}\left| \sum_{n=0}^N\frac{i^n}{n!}\int \hat f(q)\hat
g(p-q){\inn{p}{\theta q}}^n\,dq- (\hat f\circledast\hat
g)(p)\right|<\epsilon
$$
for any $N>N_\epsilon$, which completes the proof.
\end{proof}

\begin{theorem}\label{T4}
 If $f,g \in
S^\beta_\alpha(\oR^d)$, where $\beta<1/2$, then series~\eqref{*}
is absolutely summable in the space $S^\beta_\alpha(\oR^d)$, and
its sum is the function $f\times g$ defined by~\eqref{1.5}.
\end{theorem}

\begin{proof} We note that if $\beta<1/2$ and the space
$S^\beta_\alpha$ is nontrivial, then   $\alpha>\beta$. As before,
we  let $h_n$  denote the $n$th term in series~\eqref{*}. It
suffices to show that this series is absolutely summable in the
Banach space $S^{\beta, B}_{\alpha, A}$ if $A$ and $B$ are
sufficiently large. In other words, the convergence of the number
series $\sum_n\|h_n\|_{A,B}$ should be examined.
 Let $f\in S^{\beta, B_1}_{\alpha, A_1}$ and $g\in
S^{\beta, B_2}_{\alpha, A_2}$. Then we have
\begin{equation}
 |\partial^\kappa f(x)|\leq
 \|f\|_{A_1,B_1}B_1^{|\kappa|}\kappa^{\beta \kappa}
 e^{-\left|x/A_1\right|^{1/\alpha}},\quad
|\partial^\kappa g(x)|\leq
 \|g\|_{A_2,B_2}B_2^{|\kappa|} \kappa^{\beta \kappa}
 e^{-\left|x/A_2\right|^{1/\alpha}}.
 \label{3.3}
\end{equation}
We let $\mu$ and $\nu$ denote  the multi-indices in $\oZ_+^d$ that
correspond to the $n$-tuples  $(\mu_1,\dots,\mu_n)$ and
$(\nu_1,\dots,\nu_n)$ involved in~\eqref{*}. These multi-indices
are determined by the equations
$\partial^\mu=\partial_{\mu_1}\dots
\partial_{\mu_n}$, $\partial^\nu=\partial_{\nu_1}\dots
\partial_{\nu_n}$. Clearly, $|\mu|=|\nu|=n$. Let $A$ be so large that
 $A^{-1/\alpha}\leq A_1^{-1/\alpha}+ A_2^{-1/\alpha}$, and let
$C=\|f\|_{A_1,B_1} \|g\|_{A_2,B_2}$. Using Leibniz's formula and
the elementary inequalities  $(l+m)^{l+m}\leq e^{l+m}l^lm^m$ and
$l^lm^m\leq (l+m)^{l+m}$, we obtain
\begin{multline*}
e^{\left|x/A\right|^{1/\alpha}} |\partial^\kappa(\partial^\mu
f\partial^\nu g)(x)| \leq \\ \leq C\sum_\lambda
\binom{\kappa}\lambda
B_1^{|\kappa-\lambda+\mu|}B_2^{|\lambda+\nu|}
(\kappa-\lambda+\mu)^{\beta (\kappa-\lambda+\mu)}
(\lambda+\nu)^{\beta(\lambda+\nu)}\leq \\
\leq C B_1^{|\mu|}B_2^{|\nu|}e^{\beta|\kappa+\mu+\nu|}\mu^{\beta
\mu}\nu^{\beta \nu} \sum_\lambda \binom{\kappa}\lambda
B_1^{|\kappa-\lambda|}B_2^{|\lambda|}
(\kappa-\lambda)^{\beta (\kappa-\lambda)}\lambda^{\beta \lambda}\leq \\
\leq C(B_1B_2e^{2\beta})^n n^{2\beta
n}[e^\beta(B_1+B_2)]^{|\kappa|}\kappa^{\beta \kappa}.
 \end{multline*}
Taking  $B\geq e^\beta(B_1+B_2)$,  we obtain the estimate
\begin{equation}
\|h_n\|_{A,B}\leq C(B_1B_2e^{2\beta}|\theta|)^n\,\frac{n^{2\beta
n}}{n!}.
 \label{3.4}
\end{equation}
Using the inequality $n!\geq n^n/e^n$, we deduce that the series
$\sum_n\|h_n\|_{A,B}$ is indeed convergent under the condition
$\beta<1/2$. Now, we take into account that the Fourier
transformation is a topological isomorphism of
$S^\beta_\alpha(\oR^d)$ onto $S_\beta^\alpha(\oR^d)$ and apply
Theorem~\ref{T3}, which shows that the function $f\times g$ is the
sum of
 absolutely summable series~\eqref{*}. Theorem~4 is thus
proved.
\end{proof}

\begin{corollary}
 The  twisted product $\times$ on the Schwartz space
 $S(\oR^d)$ $($as well as on any space
$S^\beta_\alpha(\oR^d)$, where $\alpha\ge\beta\ge 1/2$$)$ is a
continuous extension of  $\star$-product~\eqref{*} of the
topological algebras $S^{\beta'}_\alpha(\oR^d)$, $\beta'<1/2$, for
which product~\eqref{*} is well defined.
\end{corollary}

Indeed, any nontrivial space $S^{\beta'}_\alpha(\oR^d)$ is dense
in $S(\oR^d)$ and also in $S^\beta_\alpha(\oR^d)$, where
$\beta>\beta'$. Therefore, $(f,g)\to f\times g$ is a unique
continuous map $S(\oR^d)\times S(\oR^d)\to S(\oR^d)$ (and
$S^\beta_\alpha(\oR^d)\times S^\beta_\alpha(\oR^d)\to
S^\beta_\alpha(\oR^d)$)  coinciding with the map $(f,g)\to f\star
g$ on $S^{\beta'}_\alpha(\oR^d)\times S^{\beta'}_\alpha(\oR^d)$.

\section{Continuity of the deformation}
\label{S5}
 We  now show that if $\theta \to 0$, then the product $f \times_\theta g$
tends to the ordinary product  $f\cdot g$ in the topology of the
algebras containing these functions.

\begin{theorem}\label{T5}
Let $f,g\in S^\beta_\alpha(\oR^d)$, where $\alpha\ge \beta$. The
product $f\times_\theta g$ depends continuously on the
noncommutativity parameter $\theta$.
\end{theorem}

\begin{proof}
Decomposition~\eqref{2.4} reduces the problem to
verifying that the operator on $S_\beta^\alpha(\oR^{2d})$
consisting in multiplication  by $e^{-i(q,\theta p)}$ is
continuous in the parameter $\theta$. It suffices to show this for
 $\theta=0$. We use the notation $s=(p,q)$ and $e_\theta(s)=e^{-i(q,\theta p)}$.
The analysis performed in Sec.~3 shows that
\begin{equation}
|\partial^\kappa(1-e_\theta(s))|\leq C_\epsilon
A_\epsilon^{|\kappa|} \kappa^{\alpha \kappa} e^{|\epsilon
s|^{1/\beta}}
 \label{4.1}
\end{equation}
for any $\epsilon>0$ and this estimate is uniform in  $\theta$ for
$|\theta|\le 1$. (If $\theta$ is bounded thus, then we can set
$|\theta|=1$ in~\eqref{2.7}.) Furthermore, using the Taylor series
expansion, we see that $1-e_\theta(s)=|\theta|\,\chi_\theta(s)$,
where $|\chi_\theta(s)|\le e^{|s|^2}$ for  $|\theta|\le 1$. To
estimate the derivatives  of the entire function $\chi_\theta$, we
use formula~\eqref{2.5}, but we now take the radiuses  $r_j$ of
the polydisk $D_r$ to be $\sqrt{\kappa_j}$. Then we obtain
$$
|\partial^\kappa \chi_\theta(s)|\le
\frac{\kappa!}{r^\kappa}e^{2r^2+2|s|^2}\le e^{2|\kappa|}
\kappa^{\kappa/2}e^{2|s|^2}.
$$
(we use the inequality $k!\le e^d\kappa^\kappa/2^{|\kappa|}$ in
the last step). Because $S_\beta^\alpha$ is nontrivial only if
$\alpha+\beta\ge 1$, the condition $\alpha\ge\beta$ implies that
$\alpha\ge 1/2$. Therefore, we have the inequalities
\begin{equation}
|\partial^\kappa(1-e_\theta(s))|\leq |\theta|e^{2|\kappa|}
\kappa^{\alpha\kappa}e^{2|s|^2}.
 \label{4.2}
\end{equation}
in addition to~\eqref{4.1}. Let $h\in S^{\alpha, A}_{\beta,
B}(\oR^{2d})$. We  show that there are constants $A'\ge A,B'\ge B$
such that $\|(1-e_\theta)h\|_{A', B'}\to 0$ as $|\theta|\to 0$. To
simplify the formulas in what follows, we set $B=1/3^\beta$
without loss of generality, because $S_\beta^\alpha$ is invariant
under
 dilations and
 $e_{\theta}(\lambda s)=e_{\lambda^2 \theta}(s)$. Then
\begin{equation}
|\partial^\kappa h(s)|\leq
\|h\|_{A,B}A^{|\kappa|}\kappa^{\alpha\kappa}e^{-3|s|^{1/\beta}}.
 \label{4.3}
\end{equation}
Applying Leibniz's formula and using inequality~\eqref{4.1} with
$\epsilon=1$ and inequalities  \eqref{4.2} and \eqref{4.3}, we
obtain the two estimates
$$
|\partial^\kappa[(1-e_\theta(s)) h(s)]|\leq\begin{cases} C_h
(A+A_1)^{|\kappa|}\kappa^{\alpha\kappa}e^{-2|s|^{1/\beta}},
\\
|\theta|C'_h(A+e^2)^{|\kappa|}\kappa^{\alpha\kappa}e^{2|s|^2}.
\end{cases}
 $$
Let $A'= A+\max(A_1,e^2)$ and $B'= 1$. Then we have
$$
\sup_\kappa
e^{|s/B'|^{1/\beta}}\frac{|\partial^\kappa[(1-e_\theta(s))
h(s)]|}{A'^{|\kappa|} \kappa^{\alpha\kappa}}\leq\begin{cases} C_h
e^{-|R|^{1/\beta}},&
|s|\ge R,\\
|\theta|C'_h e^{2|R|^2+|R|^{1/\beta}},& |s|<R.
\end{cases}
 $$
Given $\delta>0$, we choose  $R$ such that $C_h
e^{-|R|^{1/\beta}}\le \delta$.  Then $\|(1-e_\theta)h\|_{A',
B'}\le \delta$ for $|\theta|\le (\delta/C'_h)
e^{-2|R|^2-|R|^{1/\beta}}$. The theorem is proved.
\end{proof}

\section{Conclusion}
\label{S6}
 The performed analysis  shows that the spaces  of analytic
test functions that were previously used  to construct a quantum
theory of nonlocal interactions~\cite{FS,E2,S99} are topological
algebras under the star product. This means that they can also be
used in  QFT on a noncommutative space-time along with the
functional analytic methods developed in extending Wightman's
axiomatic approach to nonlocal fields.

Some authors (see, e.g.,~\cite{Ch,FW,Mof}) considered a
$\star$-product of field operators  $\phi(x)$ at different
space-time points, using the  definition
\begin{equation}
\phi(x_1)\star
\phi(x_2)=e^{(i/2)\,\theta^{\mu\nu}(\partial/\partial x_1^\mu)
(\partial/\partial x_2^\nu)}\phi(x_1)\phi(x_2).
 \label{5.1}
\end{equation}
This definition can easily be extended to any finite number of
operators at different points (formula (2.24) in~\cite{Sz}). The
axiomatic formulation of noncommutative QFT proposed in~\cite{Ch}
is based on the corresponding modification of the Wightman
functions written as the vacuum expectation value
\begin{equation}
\langle
0|\phi(x_1)\star\phi(x_2)\star\dots\star\phi(x_n)|0\rangle.
 \label{5.2}
\end{equation}
 There is only one way to give a rigorous mathematical meaning to
 formal definitions~\eqref{5.1} and \eqref{5.2}. Namely, the
infinite-order differential operator in~\eqref{5.1} should be
regarded as the dual of the operator
$e^{(i/2)\,\theta^{\mu\nu}(\partial/\partial x_1^\mu)
(\partial/\partial x_2^\nu)}$ acting on suitable test functions.
Clearly, the latter operator is the Fourier transform of the
multiplier $e^{-i\inn{p_1}{\theta p_2}}$ and is well defined on
the spaces $S^\beta_\alpha(\oR^{2d})$ whose indices satisfy the
restriction $\alpha\ge\beta$ established by Theorem~\ref{T1}. The
arguments used to prove Theorem~\ref{T4} show that under the
stronger condition $\beta<1/2$, the series expansion of this
operator converges on every test function. Such test function
spaces can be used as a natural initial domain of this operator
with a possible further extension  depending on the model under
consideration.

In conclusion we note that in  developing  the
Weyl-Wigner-Groenewold-Moyal approach to quantum mechanics,  much
attention was given to specifying those pairs  of tempered
distributions whose   twisted product can be formed,
see~\cite{G-BV}. The motivation for this extension is obvious
because it is desirable to include  as many physical observables
in the formalism as possible. The analysis performed here allows
constructing larger $\star$-algebras of generalized functions
including ultradistributions and hyperfunctions. This construction
will be detailed in a subsequent paper.

\section*{Appendix}

The following simple lemma is useful in examining
product~\eqref{1.4}  and in finding the conditions under which
series~\eqref{*} converges in the spaces $S^\beta_\alpha$.

\begin{lemma}
If the space $S^\beta_\alpha(\oR)$ is nontrivial, then it contains
a function $f$ such that
$$
|\partial^n f(0)|\ge  n^{\beta n} \eqno{(A1)}
$$
for all $n=0,1,2,\dots$, and the space $S_\beta^\alpha(\oR)$
contains an even nonnegative function $\hat g$ satisfying the
inequality
$$
\hat g(s)\ge  e^{-|s|^{1/\beta}}.
\eqno{(A2)}
$$
\end{lemma}

\begin{proof}
We note that the first statement of the lemma follows from the
second. Indeed, let $g={\mathcal F}^{-1}(\hat g)$. Clearly,
$\partial^n g(0)=0$ for every odd $n$, and
$$
|\partial^n  g(0)|= \frac{1}{2\pi}\int s^n \hat g(s)ds\ge
\frac{1}{2\pi} \int s^n e^{-|s|^{1/\beta}}ds
$$
for every even $n$. The maximum of the last integrand occurs when
$s=(\beta n)^\beta$. We let $s_n$ denote this number. If
$\beta>1$, then the function $|s|^{1/\beta}$  is subadditive, and
we have the inequalities
$$
\int_{s_n}^{s_n+1} s^n e^{-|s|^{1/\beta}}ds\ge (s_n+1)^n
e^{-(s_n+1)^{1/\beta}} \ge (\beta n)^{\beta n}e^{-\beta n-1}.
$$
A function $f(t)$ with  property (A1) is obtainable from
$g(t)+g'(t)$ by an appropriate scaling transformation. If
$0<\beta<1$, then we  use the inequality $|s+\sigma|^{1/\beta}\leq
2^{1/\beta}\bigl(|s|^{1/\beta}+|\sigma|^{1/\beta}\bigr)$ instead
of subadditivity, which slightly complicates the formulas but
yields the same result.

We now show that  there exists a function $\hat g\in
S_\beta^\alpha$ satisfying condition $(A2)$. For simplicity, let
$\beta>1$ as before. We use the fact that the space
$S_\beta^\alpha$ is an algebra under  (ordinary) multiplication
and is translation and dilation invariant. Starting from any
nontrivial element in it and applying these operations, we can
construct an even nonnegative function $\omega$ that also belongs
to $S_\beta^\alpha$ and has the properties
  $$ |\partial^\kappa \omega(s)|\le CA^\kappa \kappa^{\alpha \kappa}
 e^{-2|s|^{1/\beta}},\qquad\int_{-1}^{+1}\omega(s)ds=e,
 $$
 where  $C, A>0$ are sufficiently large  constants.
  We set
$$
\hat g(s)=\int e^{-|s-\sigma|^{1/\beta}}\omega(\sigma)d\sigma.
$$
Clearly, $\hat g$ is also an even nonnegative function and belongs
to $S_\beta^\alpha$. Indeed, using the subadditivity  of
$|s|^{1/\beta}$, we obtain
$$ |\partial^\kappa \hat g(s)|\le \int
e^{-|\sigma|^{1/\beta}}|\partial^\kappa
\omega(s-\sigma)|\,d\sigma\le C'A^\kappa \kappa^{\alpha \kappa}
 e^{-|s|^{1/\beta}},
$$
 where $C'=C\int e^{-|\sigma|^{1/\beta}}d\sigma$. Furthermore,  $\hat g(s)$
satisfies the  lower bound
$$
\hat g(s)\ge e^{-(|s|+1|)^{1/\beta}}
\int_{-1}^{+1}\omega(\sigma)d\sigma\ge  e^{-|s|^{1/\beta}}.
$$
This completes the proof.
\end{proof}

\begin{acknowledgements}
This paper was supported  by the Russian Foundation for Basic
Research (Grant No.~05-01-01049) and the Program for Supporting
Leading Scientific Schools (Grant No.~LSS-4401.2006.2).
\end{acknowledgements}

\end{document}